\documentclass{aastex}
\usepackage{emulateapj5}
\usepackage{apjfonts}

\gdef\g568{L451}
\gdef\h50min{$h_{50}^{-1}$}
\gdef\h75{$h_{75}^{-1}$}
\gdef\kms{km\,s$^{-1}$}
\gdef\j0848{RX\,J0848+4453}
\gdef\3727{[O\,{\sc ii}]\,3727\,\AA}
\gdef\5007{[O\,{\sc iii}]\,5007\,\AA}
\gdef\oii{[O\,{\sc ii}]}
\gdef\cl1{RDCS J0848+4453}
\shorttitle{A Massive Disk Galaxy at $z=1.34$}
\slugcomment{Accepted for publication in the Astrophysical Journal Letters}
\begin{document}

\title{A Massive Disk Galaxy at $z=1.34$
\altaffilmark{1,2}}

\author{Pieter G. van Dokkum\altaffilmark{3}}\affil{California Institute
of Technology, MS105-24, Pasadena, CA 91125}
\author{\and S. A. Stanford\altaffilmark{4}}\affil{Institute of Geophysics
and Planetary Physics, Lawrence Livermore National Laboratory}

\altaffiltext{1}{Based on observations with the NASA/ESA {\em Hubble Space
Telescope}, obtained at the Space Telescope Science Institute, which
is operated by AURA, Inc., under NASA contract NAS 5--26555.}
\altaffiltext{2}
{Based on observations obtained at the W.\ M.\ Keck Observatory,
which is operated jointly by the California Institute of
Technology and the University of California.}
\altaffiltext{3}{Hubble Fellow}

\altaffiltext{4}{Physics Department, University of California-Davis, Davis, CA 
95616, USA}

\begin{abstract}

We report the discovery
of a rapidly rotating disk in a $K$-selected galaxy at $z=1.34$.
Spatially resolved kinematics are determined from
deep, moderate resolution Keck spectroscopy covering the
redshifted \3727{} line.
A conservative estimate of the maximum
rotation velocity $V_{\rm max} = 290 \pm 20$\,\kms, placing the galaxy
among the most rapid rotators known. A
{\em Hubble Space Telescope} WFPC2 image taken in the
$I_{F814W}$ filter (rest-frame $U$) shows that the \oii\ emission
originates in a ring of star forming regions $\sim 3''$ ($\sim
20$\,\h75\,kpc) across.
In the $K$-band the galaxy is compact, and its position
coincides with the center of the star forming ring seen in the
rest-frame near-UV. The galaxy has
$M_V = -22.4 \pm 0.2$ ($\sim 3 L_*$) in the rest-frame,
and its spectral energy distribution is very well fitted by that of
a redshifted Sb/c galaxy.
The observed kinematics, morphology and
spectral energy distribution are consistent with a massive bulge-disk
system at $z=1.34$.
The lower limit on $V_{\rm max}$ gives an upper limit on the
offset from the present-day Tully-Fisher relation.
The galaxy is
overluminous by less than $0.7 \pm 0.4$ magnitudes in rest-frame
$V$, consistent with
previous studies of disk galaxies at lower redshift.
Taken together,
these results suggest that some of the most luminous
spiral galaxies in the nearby Universe were already in place
$\sim 10^{10}$\,yr ago, placing a constraint on models for
their formation.

\end{abstract}

\keywords{
galaxies: formation ---
galaxies: evolution ---
galaxies: structure of ---
galaxies: kinematics and dynamics ---
galaxies: spiral
}

\section{Introduction}

Hierarchical galaxy formation models predict that present-day
massive disks were formed at $z\lesssim 1$ (Mo, Mao, \&
White 1998). The observational
evidence concerning the existence of large,
massive disks at high redshift is currently
inconclusive.  Studies of quasar
absorption lines have demonstrated that structures with velocity
widths up to $\sim 200$\,\kms\ exist out to $z\sim 4$ (e.g.,
Prochaska \& Wolfe 1997), but these may
result from the motions of distinct kinematic systems within a common
halo (e.g., Maller et al.\ 1999, Wolfe \& Prochaska 2000).
Another complication in kinematic studies of star forming galaxies
at $z>2$ is the effect of outflows on the observed line widths
(Franx et al.\ 1997; Pettini et al.\ 2001).

Parallel to these studies there has been significant progress in the study of
``normal'', morphologically identified disk galaxies at intermediate
redshift.
Studies of the structure of distant field galaxies
with the {\em Hubble Space Telescope} (HST) indicate
that the sizes of disks do not show strong evolution to
$z\sim 1$, but that their rest-frame
$B$ band surface brightnesses may be higher by $\sim 1$ magnitude
at that redshift (Lilly et al.\ 1998; Simard et al.\ 1999).
The interpretation is hampered by selection effects (see Simard et
al.\ 1999), and the implicit assumption that structural and
surface brightness evolution of galaxies can be separated
(see, e.g., Mao, Mo, \& White 1998).

Vogt et al.\ (1996, 1997) provided an important additional
constraint by measuring spatially resolved kinematics of disk galaxies
out to $z\sim 1$, thus allowing comparisons of sizes and
luminosities of galaxies at fixed rotation velocity.
Current data suggest that large, rapidly rotating
disks exist at least out to $z\sim 1$, and that their luminosities
are at most slightly brighter than equally massive disks today
(Vogt et al.\ 1997).

In this {\em Letter}, we report on  observations of
a rapidly rotating, luminous galaxy at $z=1.34$ dubbed \g568.
This galaxy shares
many of its properties with the most massive spiral galaxies
in the local Universe, and provides direct information on
the properties of massive disks in the hitherto virtually
unexplored redshift range $1 - 2$. We use
$H_0 = 75$\,\kms\,Mpc$^{-1}$, $\Omega_m = 0.3$ and $\Omega_{\Lambda}=0$.

\section{Observations}

Galaxy \g568{} ($\alpha = 8^d 48^m 40.0^s$,
$\delta = +44^{\circ} 54' 13''$; J2000)
is located in the vicinity of the cluster \cl1\ at $z=1.27$
(Stanford et al.\ 1997). This field has
been studied extensively, both from the ground and with HST.  Ground
based $BRIJK_s$ imaging data were previously reported in Stanford et
al.\ (1997). Deep HST WFPC2 imaging data were described in van Dokkum et al.\
(2001a).  Ten exposures were obtained in the $I_{F814W}$ filter for a
total of 27.8\,ks. The galaxy falls just outside the region
covered by a NICMOS $H_{F160W}$ mosaic.

We included \g568{} in a multi-slit spectroscopic observation of \cl1.
The galaxy has $K_s = 18.2$ (Stanford et al.\ 1997) and $z_{\rm phot}
= 1.40 \pm 0.08$ (see van Dokkum et al.\ 2001a), and was selected
for deep spectroscopy because of the proximity of its photometric
redshift to the redshift of the cluster.
The field was observed on 2001 January 20--21 with the Low
Resolution Imaging Spectrograph (Oke et al.\ 1995) on the Keck II
Telescope, using multi-slit masks. The 600\,l\,mm$^{-1}$ grating
blazed at 1\,$\mu$m was used with $1\farcs 2$ wide slits, giving
a velocity resolution of $\sigma_{\rm instr} \approx 80$\,\kms\
at 9000\,\AA. Conditions
were photometric, and the seeing was $\approx 0\farcs 9$. Between
exposures the spectra were moved along the slit to facilitate sky
subtraction.  The total integration time was 28.8\,ks.  The data
reduction followed standard procedures for dithered multi-slit
spectroscopic data (see, e.g., Stanford et al.\ 1997); details of the
reduction are described in P.\ G.\ van Dokkum and S.\ A.\ Stanford, in
preparation.

\section{Analysis}

\subsection{Redshift}

The spectrum of \g568\ is shown in
Fig.\ \ref{spec.plot}.
The spectrum shows a prominent emission line at 8703\,{\AA} and
several absorption lines.
The line is a doublet identified
as \3727{} at $z=1.335$, and we conclude that \g568{} is a field galaxy
unrelated to the foreground cluster at $z=1.27$. The spectroscopic
redshift is within $1 \sigma$ of the photometric redshift.

\vbox{
\begin{center}
\leavevmode
\hbox{%
\epsfxsize=8.2cm
\epsffile{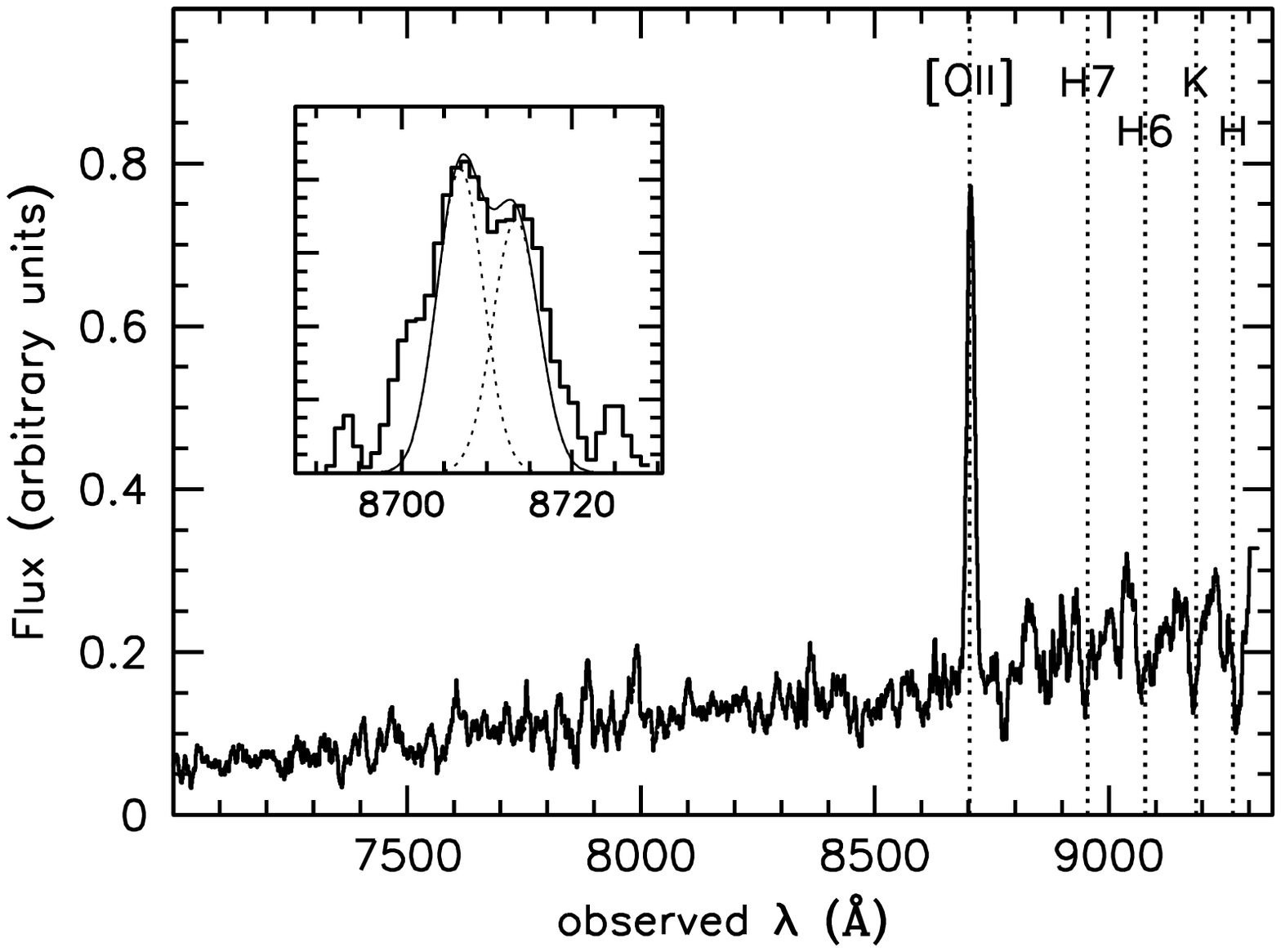}}
\figcaption{\small
Integrated spectrum of \g568, smoothed by a 13\,\AA{}
wide boxcar filter. The \3727\
line redshifted to $z=1.335$ can be identified, as well as higher
order Balmer lines and the Ca\,{\sc ii} H,K lines.
The inset shows the (unsmoothed)
\oii\ doublet in the region of highest S/N ratio,
which is $\sim 200$\,\kms removed from the systemic velocity.
\label{spec.plot}}
\end{center}}

\subsection{Morphology and Spectral Energy Distribution}

The HST $I_{F814W}$ image of \g568\ is shown in Fig.\ \ref{hst.plot}.
The galaxy is elongated, and has an irregular appearance.
Several individual concentrations
can be identified. Furthermore, the
light is not concentrated towards the center but appears
to be distributed in a ring.
These features may indicate that the galaxy is
not a relaxed system, and experienced a merger or interaction.
However, since the observed $I$-band corresponds to the rest-frame
$U$-band for a galaxy at $z=1.34$, the irregular morphology
can also be due to patchy star formation.

In Fig.\ \ref{contour.plot}
the morphology in the $I$ band is compared
to that in the $K$-band.
For an object at $z=1.34$ the $K$ band corresponds
to the $z$-band in the rest-frame,
which is much less sensitive to ongoing star
formation than the $U$-band.
The resolution of the $K$-band image
is slightly enhanced (from $1\farcs 3$ to $0\farcs 9$ FWHM)
by a {\sc clean}
reconstruction (H\"ogbom 1974) 
using a nearby bright star.
The $I$-band image is created by smoothing the
WFPC2 image to $0\farcs 9$ resolution and resampling.
The galaxy is compact in $K$, and its center coincides
with the center of the ring seen in the $I$-band. The $K$-band
morphology argues against the merger hypothesis, and
we conclude that the irregular appearance of \g568{} in the
rest-frame near-UV is probably caused by star formation in
an otherwise regular galaxy.

\vbox{
\begin{center}
\leavevmode
\hbox{%
\epsfxsize=8.2cm
\epsffile{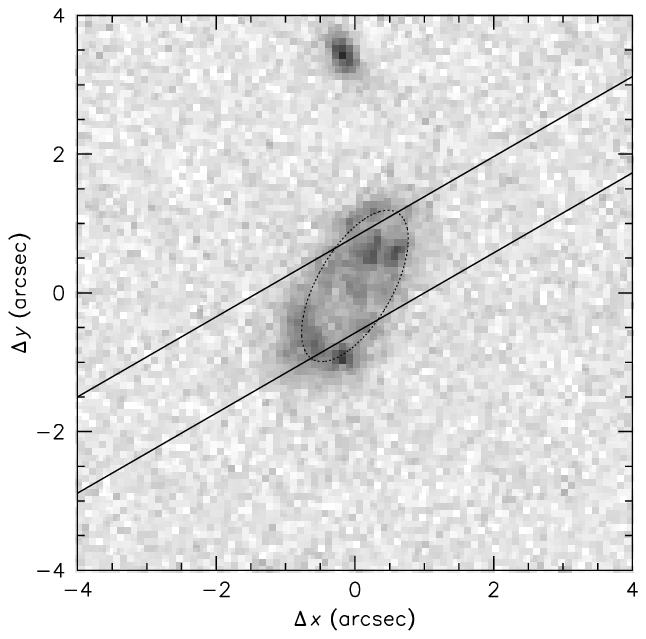}}
\figcaption{\small
HST WFPC2 image of \g568\ in the $I_{F814W}$ filter.
The solid lines show the slit that was used
in the Keck observations. The ellipse indicates the adopted inclination
and position angle of \g568. The galaxy has an irregular morphology
in the rest-frame $U$ band.
\label{hst.plot}}
\end{center}}

\vbox{
\begin{center}
\leavevmode
\hbox{%
\epsfxsize=8.2cm
\epsffile{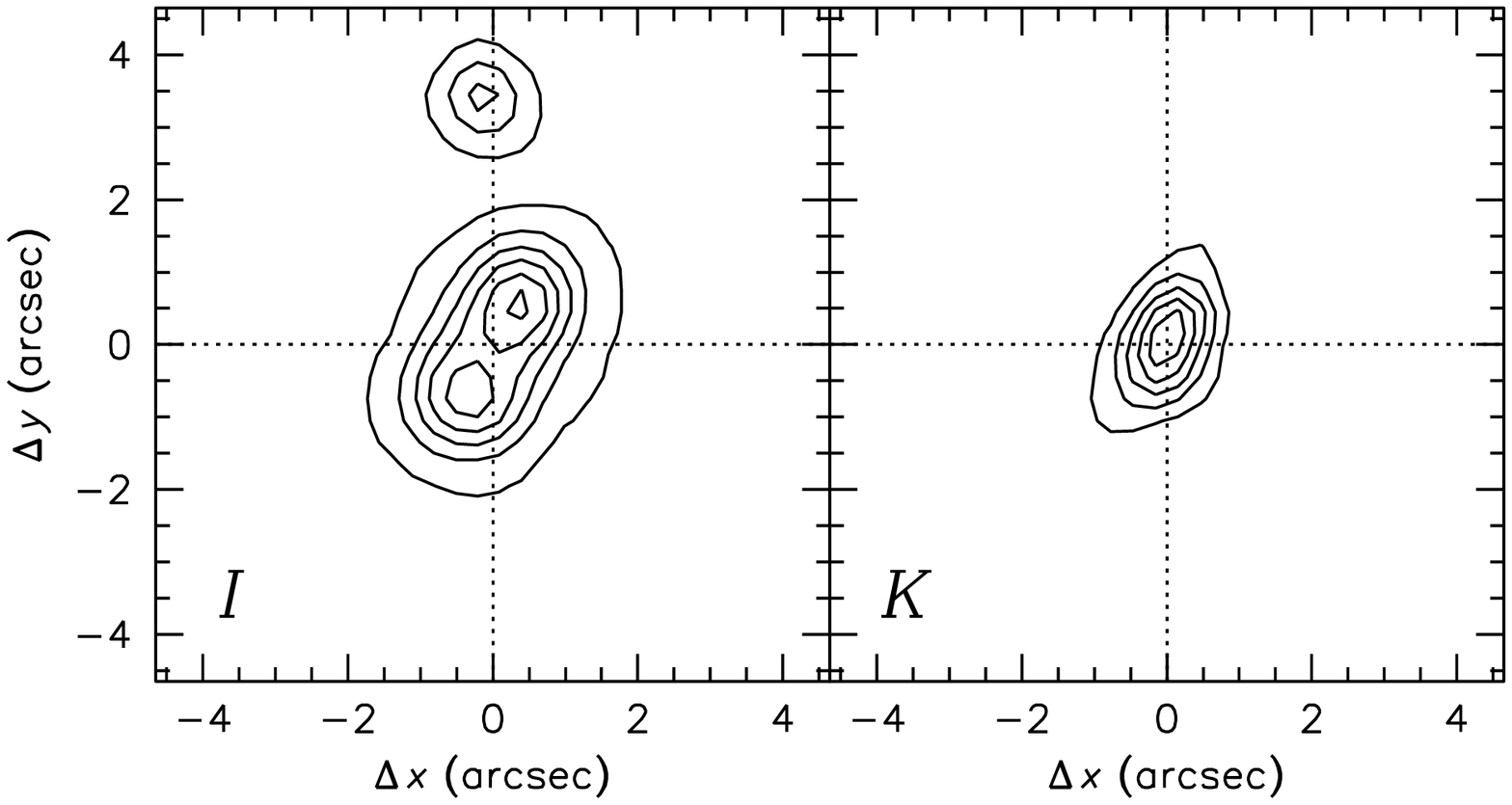}}
\figcaption{\small
Comparison of morphology in $I_{F814W}$ and $K$ filters (rest-frame
$U$ and $z$). The $I_{F814W}$ image was smoothed and resampled to
match the $K$ band resolution. Note the regular, compact morphology
of the $K$ band image in comparison to the smoothed $I$
band image.
\label{contour.plot}}
\end{center}}

This interpretation is supported by the integrated
spectral energy distribution
(SED) of \g568, shown in Fig.\ \ref{sed.plot}. The galaxy is quite
red: $I-K=3.8$ in a $3''$ aperture, implying that
the $K$-band morphology represents the distribution
of the bulk of the stellar mass. Remarkably,
the SED is very well fitted by
that of {\em present-day} Sb/c galaxies as obtained from Coleman,
Wu, \& Weedman (1980).
We conclude that \g568\ is probably not a merger, but
a very luminous early-type spiral galaxy.
We use the observed $J$ magnitude and $J-K$ color
to obtain the rest-frame $V$ magnitude (see van Dokkum et al.\ 1998).
Using $J=20.4$ and $J-K_s = 2.25$ we obtain
$M_V = -22.3 \pm 0.2$, or $\sim 3 L_*$.

Early-type spiral galaxies in
the local Universe often show a similar
wavelength dependence of
morphology. The near-UV light traces star
forming regions which may be distributed in a disk or a ring,
whereas the near-IR light is usually dominated
by the bulge (e.g., Kuchinski et al.\ 2000).

\vbox{
\begin{center}
\leavevmode
\hbox{%
\epsfxsize=8cm
\epsffile{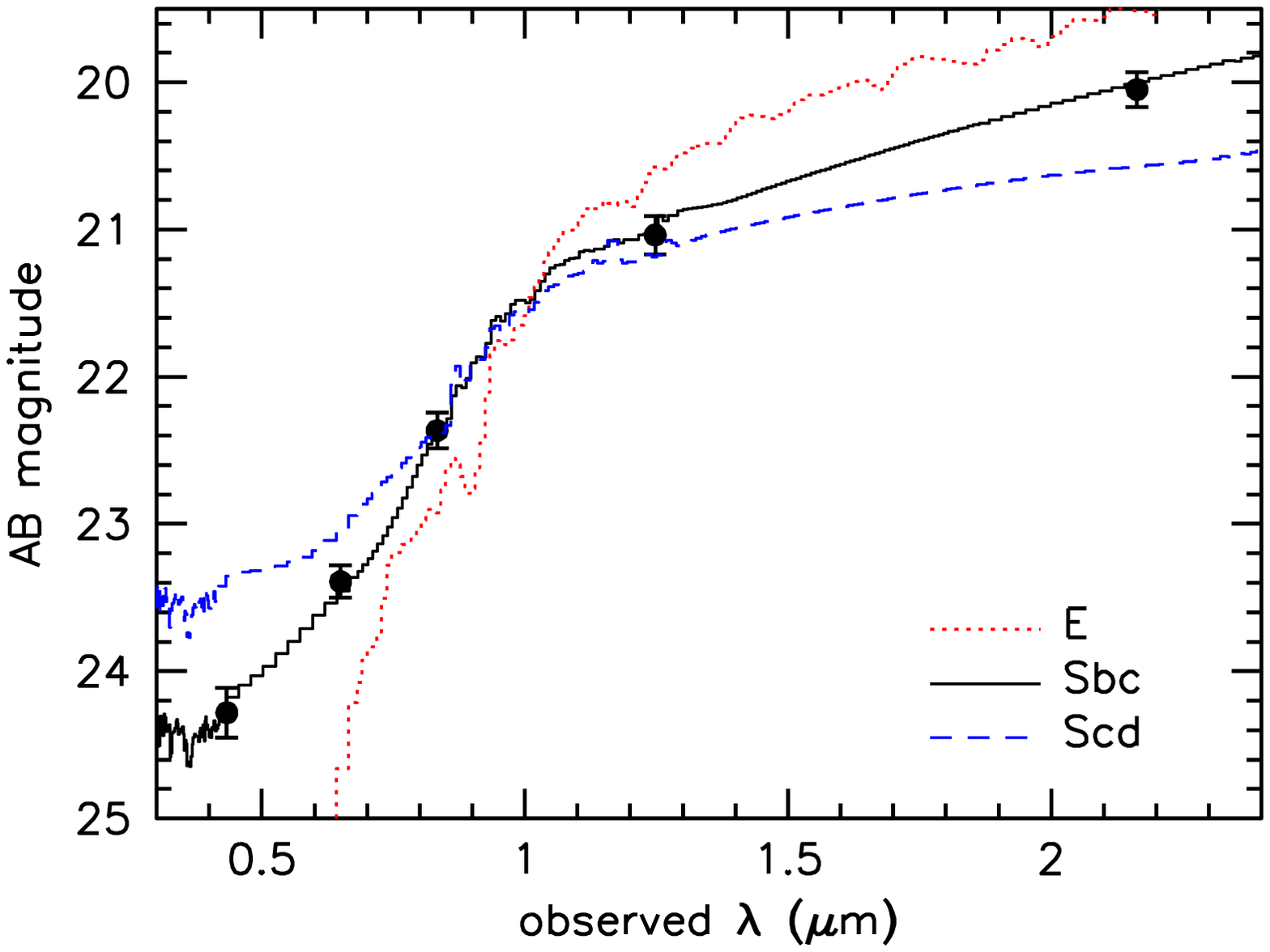}}
\figcaption{\small
Spectral energy distribution of \g568. Overplotted are the SEDs of
various galaxy types in the nearby Universe (Coleman, Wu, \& Weedman
1980), redshifted to $z=1.335$.
The observed SED is much redder than that of star forming late-type
galaxies, and is very well fitted by that of present-day Sb/c
galaxies.
\label{sed.plot}}
\end{center}}

\subsection{Inclination}

The resolution and depth of our ground based data are not sufficient to
determine the size and inclination of the disk. Therefore, we
determine these parameters from the $I_{F814W}$ image, with
the caveat that the light is dominated by star forming regions.
The position angle and inclination of the galaxy are obtained
from an ellipse fit to the star forming ring (indicated in Fig.\
\ref{hst.plot}). We find PA$=150 \pm 7$ degrees
and $(b/a)=0.44 \pm 0.05$. Assuming that \g568\ has an
intrinsically circular disk whose thickness is less than $\sim 20$\,\%
of its diameter, we infer that \g568\ is inclined
by $65 \pm 5$ degrees.
The diameter of the ring is $\approx 2\farcs 7$, corresponding to
$\approx 19$\,\h75\,kpc.
High resolution
images in the rest-frame optical are needed to determine
the exponential scale length of the disk.
Such data may also be used to determine
whether the galaxy has a bar, as suggested
by its ring-like morphology in the rest-frame near-UV
(e.g., Kormendy 1979).

\subsection{Rotation Velocity and Tully-Fisher Relation}

The position of the slit that was used in the LRIS observations is
indicated in Fig.\ \ref{hst.plot}. The major axis of the galaxy and the
slit are misaligned by only $\sim 30^{\circ}$.
As is evident in Fig.\ \ref{spec2D.plot} the \oii\ line displays a
significant velocity gradient. Radial velocities at each position
along the slit are obtained by fitting a double Gaussian profile.
The line ratio $r_{\mbox{\small O\,{\sc ii}}} \equiv
F(\lambda 3729)/F(\lambda 3726)$ is
determined from the observed profile of the doublet
at the position with the highest
signal-to-noise ratio (see Fig.\ \ref{spec.plot}),
and assumed to be constant over the extent of
the galaxy. The observed line ratio
$r_{\mbox{\small O\,{\sc ii}}} \sim 0.8$ implies an
electron density $N_e \sim 10^3$\,cm$^{-3}$ (Odell \& Castaneda 1984).
Such a high density is unusual for extragalactic
H\,{\sc ii} regions (e.g., Shields 1990);
we note that \g568\ is not detected in a 185\,ks Chandra X-ray
observation (Stanford et al.\ 2001).

The resulting velocity curve is shown in the lower panel
of Fig.\ \ref{spec2D.plot}. Horizontal bars show the rows that
were averaged in the 2D spectrum, and the positions are intensity
weighted averages. The open circle shows the velocity derived from the
Ca\,{\sc ii} H and K lines.
The data points are highly
correlated because of the seeing, and the shape of the rotation
curve is not constrained by our data.
The total velocity range is $510$\,\kms,
and the implied maximum observed rotation velocity is $255 \pm
15$ \kms.

\vbox{
\begin{center}
\leavevmode
\hbox{%
\epsfxsize=8cm
\epsffile{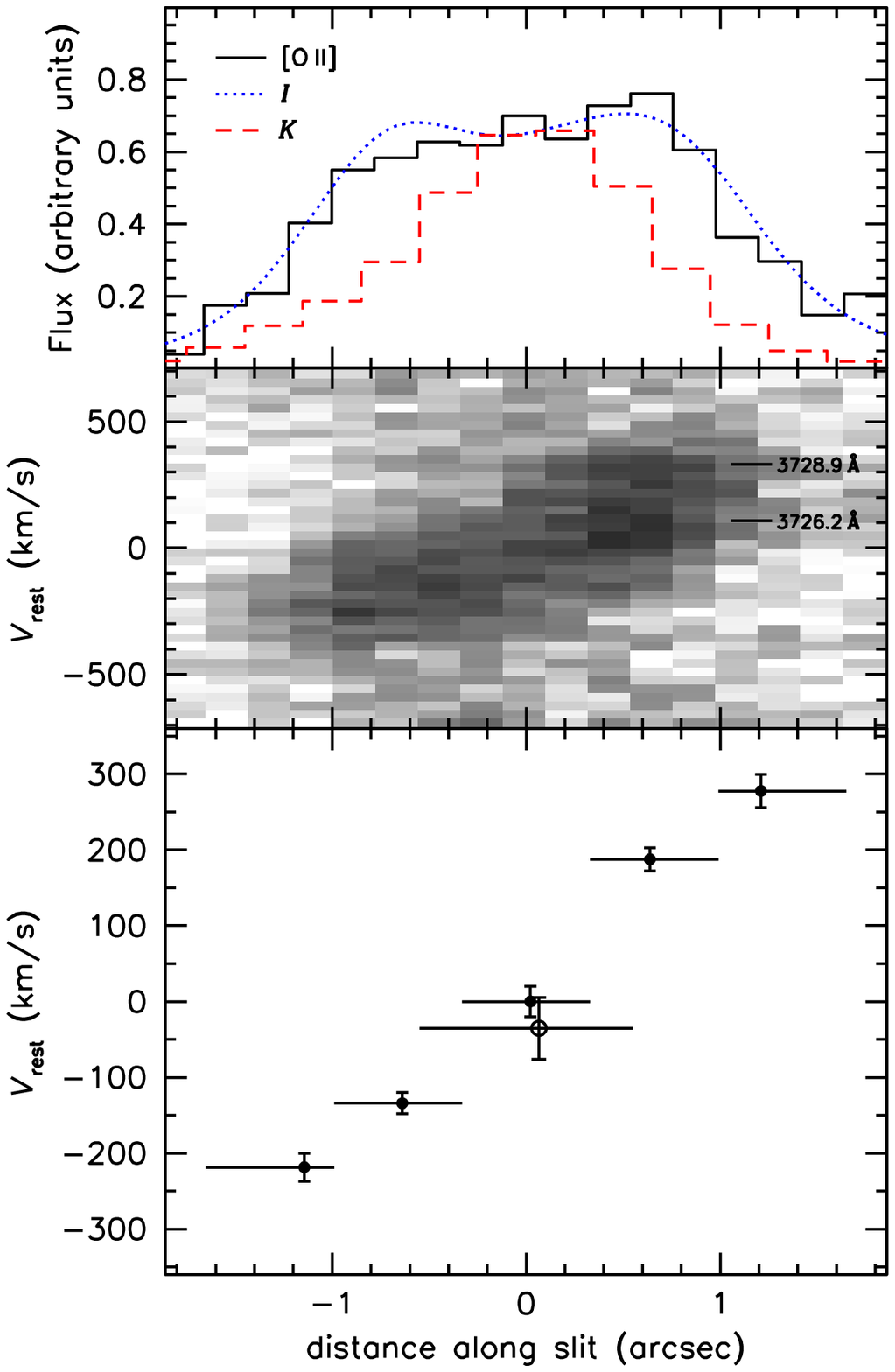}}
\figcaption{\small
Radial velocity along the slit derived from the \3727\
line (lower panel). The middle panel shows the 2D spectrum, and the top panel
shows the scaled distributions of the \oii\ line, the $I$-band light, and the
$K$-band light along the slit.
The observed velocity range $2 V_{\rm max} \approx 510$\,\kms.
The open circle shows the velocity derived from the
Ca\,{\sc ii} H and K lines.
\label{spec2D.plot}}
\end{center}}

The observed velocities are a convolution of the underlying velocity
field, the distribution of the line emitting gas, the inclination of
the galaxy, and the position of the slit.  Vogt et al.\ (1996, 1997)
used exponential disk models to interpret their observed rotation
curves. Such models are not appropriate for \g568: the top panel of
Fig.\ \ref{spec2D.plot} shows that the \oii\ emission follows
the distribution of the near-UV continuum light, which is
dominated by a few individual star forming regions (see Fig.\
\ref{hst.plot}). As a result,
modelling of the
underlying velocity field is highly degenerate.
Applying a straightforward inclination correction gives a conservative
estimate of the maximum rotation velocity $V_{\rm max} = 290 \pm
20$\,\kms.  This is strictly a lower
limit, since there is no indication that our observations sample
the flat part of the rotation curve.


Following Vogt et al.\ (1996, 1997) we determine the offset of
\g568{} from the present-day Tully-Fisher relation
(Tully \& Fisher 1977).  The
Tully-Fisher relation predicts $M_B^a = -21.4 \pm 0.2$ for $V_{\rm
max} = 290$\,\kms{} (e.g., Pierce \& Tully 1992, Verheijen 2001).
The uncertainty reflects the differences in slope and offset of
the relation obtained from different local samples. Applying the
extinction correction of Tully \& Fouque (1985) gives $M_V^a = -22.8
\pm 0.3$ for \g568.  Taking $B-V = 0.7$ for present-day massive spiral
galaxies (de Jong 1995) we conclude that \g568{} is overluminous by
$0.7 \pm 0.4$ magnitudes. Because $V_{\rm max}$ 
may be underestimated, this offset is strictly an upper limit.

Although it is hazardous to draw conclusions from a single object, our
results for \g568\ are consistent with studies by Vogt et al.\ (1996,
1997) of field galaxies at lower redshift. They are also consistent
with models of Ferreras \& Silk (2001), which predict $0 - 1$
magnitudes of luminosity brightening at $z = 1.3$.
We note that these models do not treat the evolution of the
bulge and the disk separately; if bulges evolve in the same way
as elliptical galaxies, the bulge of \g568\ is expected to be
$1 - 1.5$ magnitudes brighter than low redshift
bulges of the same mass (van Dokkum et al.\ 1998, 2001b).

\section{Discussion}

Provided that \g568\ is not exceptional, it may prove challenging to
produce such regular, massive bulge-disk systems as early as $z
\approx 1.3$ in current hierarchical galaxy formation models.  It is
generally thought that the properties of disks are closely tied to
those of the dark halos in which they formed (e.g., Fall \& Efstathiou
1980), and in hierarchical clustering models present-day large disks
were formed late, at $z \lesssim 1$ (e.g., Mo et al.\ 1998,
van den Bosch 1998). High redshift galaxies with large rotation
velocities do exist in these models, but they are
small. For a galaxy at $z=1.3$ with $V_{\rm max} = 290$\,\kms\
the predicted scale length $R_d \sim
3$\,kpc (Mao et al.\ 1998) --- approximately
one third the size of the star forming ring of
\g568. High resolution images in the
rest-frame optical or near-IR are necessary to measure the exponential
scale length of \g568, for a direct comparison to low redshift
samples.

The local space density of $L\geq 3L_*$ galaxies is only $\sim 7
\times 10^{-5}\, h_{75}^{3}$\,Mpc$^{-3}$ (Blanton et al.\ 2001), and
surveys over large volumes are required to determine the space density
of objects such as \g568.  As an example, the expected number of
$L\geq 3L_*$ galaxies between $z=1.3$ and $z=1.4$ in the $K$-selected
Cowie et al.\ (1996) sample is only $\sim 0.3$ (assuming no luminosity
evolution); hence their absense in this survey is not very
constraining. Combining large area $K$-band surveys (e.g., Daddi et
al.\ 2000) with accurate photometric redshifts may be the most
efficient way to select candidate massive disk galaxies at $1<z<1.5$.

In hierarchical models,
virtually all present-day $L \gtrsim 2.5 L_*$ galaxies have Ly-break
progenitor galaxies  (Baugh et al.\ 1998),
and  it is interesting to compare the properties of \g568\ to those of
known galaxies at $z\approx 3$. Ly-break galaxies are
compact (Giavalisco, Steidel, \& Macchetto 1996), and
recent \5007\ spectroscopy 
shows that they have modest line widths,
of $50 - 110$\,\kms\ (Pettini et al.\ 2001).
Taking these results at face value,
several mergers are required to build up
a galaxy of the size and mass of \g568.
After these mergers the remaining gas in the halo has to
settle into a disk. These processes need to be efficient, and
be completed in $\sim 2 \times 10^9$\,yr.

Alternatively, large, evolved galaxies such as \g568\ may yet have
escaped detection at higher redshift. The galaxy is quite faint
in the rest-frame ultra-violet
(see Fig.\ \ref{sed.plot}), and if it were placed at $z=3$
it would have $I_{\rm AB} \approx 26.5$, more than 1 magnitude fainter
than typical Ly-break galaxies.
Furthermore, in the region
between the star forming regions its surface brightness 
would be $I_{\rm AB} \gtrsim 29$\,arcsec$^{-2}$, undetectable even
in the Hubble Deep Fields.
The star forming regions would show as distinct
objects $\sim 2''$ and $\sim 400$\,\kms\ apart, making it very difficult
to assess the true nature of the galaxy. If
disk galaxies do not evolve strongly
in luminosity, as suggested by the Tully-Fisher
results to $z \approx 1.3$,
very deep IR imaging 
may be the only way to detect
large, massive disks at $z\sim 3$.

\acknowledgements{
We thank Richard Ellis, Wall Sargent and Rob Swaters for their
comments on the manuscript. P. G. v. D.
acknowledges support by NASA through Hubble Fellowship
grant HF-01126.01-99A
awarded by the Space Telescope Science Institute, which is
operated by the Association of Universities for Research in
Astronomy, Inc., for NASA under contract NAS 5-26555.
S. A. S. is supported by the Institute of Geophysics and Planetary Physics
(operated under the auspices of the US Department of Energy by the
UC Lawrence Livermore National Laboratory under
contract W-7405-Eng-48), and by NASA/LTSA
grant NAG5-8430.
}

\newpage


\begin{references}
\reference{}     Baugh, C. M., Cole, S., Frenk, C. S., \& Lacey, C. G. 1998,
	\apj, 498, 504
\reference{}     Blanton, M. R., et al. 2001, \aj, 121, 2358
\reference{}     Coleman, G. D., Wu, C.-C., \& Weedman, D. W. 1980,
	\apjs, 43, 393
\reference{}     Cowie, L. L., Songaila, A., Hu, E. M., \& Cohen, J. G.
	1996, \aj, 112, 839
\reference{}     Daddi, E., et al. 2000, A\&A, 361, 535
\reference{}     de Jong, R. S. 1995, Ph.D. thesis, Groningen University
\reference{}     Fall, S. M., \& Efstathiou, G. 1980, \mnras, 193, 189
\reference{}     Ferreras, I., \& Silk, J. 2001, \apj, 557, 165
\reference{}     Franx, M., Illingworth, G. D., Kelson, D. D., van Dokkum,
	P. G., \& Tran, K.-V. 1997, \apj, 486, L75
\reference{}     Giavalisco, M., Steidel, C. C., \& Macchetto, F. D. 1996,
	\apj, 470, 189
\reference{}     H\"ogbom, J. A. 1974, A\&AS, 15, 417
\reference{}     Kormendy, J. 1979, \apj, 227, 714
\reference{}     Kuchinski, L. E., et al. 2000, \apjs, 131, 441
\reference{}     Lilly, S., et al. 1998, \apj, 500, 75
\reference{}     Maller, A. H., Somerville, R. S., Prochaska, J. X.,
	\& Primack, J. R. 1999, in After the Dark Ages,
	ed. S. Holt \& E. Smith,
	AIP Press, p. 102
\reference{}     Mao, S., Mo, H. J., \& White, S. D. M. 1998,
	\mnras, 297, L71
\reference{}     Mo, H. J., Mao, S., \& White, S. D. M. 1998,
	\mnras, 295, 319
\reference{}    Odell, C. R., \& Castaneda, H. O. 1984, \apj, 283, 158
\reference{}     Oke, J. B., et al. 1995, \pasp, 107, 375
\reference{}     Pettini, M., et al.\ 2001, \apj, 554, 981
\reference{}     Pierce, M. J., \& Tully, R. B. 1992, \apj, 387, 47
\reference{}     Prochaska, J. X., \& Wolfe, A. M. 1997, \apj, 487, 73
\reference{}     Shields, G. A. 1990, ARA\&A, 28, 525
\reference{}     Simard, L., et al.  1999, \apj, 519, 563
\reference{}     Stanford, S. A., Elston, R., Eisenhardt, P. R.,
	Spinrad, H., Stern, D., \& Dey, A. 1997, \aj, 114, 2232
\reference{}     Stanford, S. A., et al.\ 2001,
	\apj, 552, 504
\reference{}     Tully, R. B., \& Fisher, J. R. 1977, A\&A, 54, 661
\reference{}     Tully, R. B., \& Fouque, P. 1985, \apjs, 58, 67
\reference{}     van den Bosch, F. C. 1998, \apj, 507, 601
\reference{}      van Dokkum, P. G., Franx, M., Kelson, D. D., \&
	Illingworth, G. D. 1998, \apj, 504, L17
\reference{}     van Dokkum, P. G., Stanford, S. A., Holden, B. P.,
	Eisenhardt, P. R., Dickinson, M., \& Elston, R. 2001a,
	\apj, 552, L101
\reference{}     van Dokkum, P. G., Franx, M., Kelson, D. D., \&
	Illingworth, G. D. 2001b, \apj, 553, L39
\reference{}     Verheijen, M. A. W. 2001, \apj, in press (astro-ph/0108225)
\reference{}     Vogt, N. P., Forbes, D. A., Phillips, A. C., Gronwall, C.,
	Faber, S. M., Illingworth, G. D., \& Koo, D. C. 1996, \apj,
	465, L15
\reference{}     Vogt, N. P., et al. 1997, \apj, 479, L121
\reference{} Wolfe, A. M., \& Prochaska, J. X. 2000,
	\apj, 545, 591
\end{references}
\end{document}